\documentclass[%
12pt,
reprint,
amsfonts,amsmath,amssymb,
aps,
prb,
]{revtex4-1}

\usepackage{mathptmx}
\usepackage{amssymb}
\usepackage{amsbsy}
\usepackage{color}
\usepackage{cancel}
\usepackage[tight]{subfigure}

\usepackage{graphicx}
\usepackage[tight]{subfigure}
\usepackage{dcolumn}
\usepackage{bm}
\usepackage[colorlinks,linkcolor={blue},citecolor={blue},urlcolor={blue}]{hyperref}
\usepackage{marginnote}

\newcommand{\ba}{\begin{array}}
\newcommand{\ea}{\end{array}}

\newcommand{\ra}{\rangle}
\newcommand{\la}{\langle}

\newcommand{\up}{\uparrow}
\newcommand{\dn}{\downarrow}
\newcommand{\nnum}{\nonumber}

\newcommand{\ket}[1]{\left\lvert#1\right\rangle}
\newcommand{\bra}[1]{\left\langle#1\right\rvert}

\begin{document}

\title{Mott transition, magnetic and orbital orders in the ground state of the two-band 
Hubbard model using variational slave-spin mean field formalism}
\author{Arun Kumar Maurya}\email{akmourya13@iisertvm.ac.in}
\author{Md.~Tahir Hossain Sarder}\email{tahir15@iisertvm.ac.in} 
\author{Amal Medhi}\email{amedhi@iisertvm.ac.in} 
\affiliation{Indian Institute of Science Education and Research Thiruvananthapuram, 
Kerala 695551, India}

\begin{abstract}
We study the ground state of the Hubbard model on a square lattice with two degenerate 
orbitals per site and at integer fillings as a function of onsite Hubbard repulsion 
$U$ and Hund's intra-atomic exchange coupling $J$. We use a variational 
slave-spin mean field (VSSMF) method which allows symmetry broken states to be studied 
within the computationally less intensive slave-spin mean field formalism, 
thus making the method more powerful to study strongly correlated electron physics.
The results show that at half-filling, the ground state at smaller
$U$ is a Slater antiferromagnet (AF) with substantial local charge fluctuations. As $U$ is
increased, the AF state develops a Heisenberg behavior, finally undergoing a first 
order transition to a Mott insulating AF state at a critical interaction $U_c$ which 
is of the order of the bandwidth. 
Introducing the Hund's coupling $J$ correlates the system more and reduces
$U_c$ drastically. At quarter-filling with one electron per site, the ground 
state at smaller $U$ is paramagnetic metallic. At finite Hund's coupling $J$, 
as interaction is increased above a lower critical value $U_{c1}$,
it goes to a fully spin polarized ferromagnetic state coexisting with
an antiferro-orbital order. The system eventually becomes Mott insulating at 
a higher critical value $U_{c2}$. The results as a function of $U$ and $J$ are 
thoroughly discussed. 
\end{abstract}
  
\maketitle
\section{Introduction}
The Hubbard model\cite{GutzwillerPhysRevLett.10.159,Kanamori_10.1143/PTP.30.275,Hubbard_doi:10.1098/rspa.1963.0204} has been used to describe various strongly correlated electron phenomena observed in several $d$ and $f$-electron
materials\cite{MottRevModPhys.40.677,Imada_RevModPhys.70.1039,Orenstein468,XiaoGrangRevModPhys.78.17}. Particularly, it has been 
extensively studied in the context of Mott metal-insulator transition (MIT)
which is obtained in transition metal oxides by varying pressure, temperature or chemical composition\cite{Imada_RevModPhys.70.1039}. MIT within the Hubbard model
has been studied using various methods such as dynamical mean field theory 
(DMFT)\cite{DMFT_RevModPhys.68.13,Zhang_PhysRevLett.70.1666,
Rozenberg_PhysRevB.49.10181,Park_PhysRevLett.101.186403, 
Rozenberg_PhysRevLett.83.3498,Bulla_PhysRevB.64.045103}, Gutzwiller variational 
theory\cite{Lu_PhysRevB.49.5687,Ferrero_PhysRevB.72.205126}, slave-boson 
mean field (SBMFT)\cite{Sigrist_EPJB2005,KotliarPhysRevB.56.12909} etc.
The picture of MIT in the one-band Hubbard model that emerges from these
studies are that at zero temperature, the frustrated model at half-filling undergoes a continuous Mott transition from a paramagnetic (PM) metallic to a paramagnetic Mott insulating state driven by electron-electron interactions. At finite temperatures below a critical value, the transition is first order in nature. The mechanism of MIT as supported by DMFT is that it occurs via shifting of spectral weight from the Fermi level to preformed Hubbard sub-bands as
the onsite Coulomb repulsion $U$ is increased, a scenario containing features of both the Brinkmann-Rice and the Mott-Hubbard mechanisms\cite{Kotliar:10.1063/1.1712502,LoganNoziers}.
For the non-frustrated model on a bipartite lattice, antiferromagnetic (AF) long range order sets in at low temperatures due to perfect Fermi surface nesting, and it preempts a Mott 
transition as the N\'{e}el temperature is higher than the MIT temperature range. The AF state becomes band insulating because of the doubling of the unit cell. 
The detailed nature of MIT depends on degrees of AF correlations and 
frustration present\cite{Chitra_PhysRevLett.83.2386}.
Qualitatively similar features of Mott transition 
were also obtained for the multi-band Hubbard model at integer 
fillings\cite{Kotliar_PhysRevB.54.R14221,Rozenberg_PhysRevB.55.R4855,Pruschke_EPJB2005,Bulla_PhysRevB.64.045103}.

The original motivation behind introducing the Hubbard model was 
to describe itinerant ferromagnetism in metals. However, the one-band model
turned out to be inadequate for the purpose. Ferromagnetism in the one-band Hubbard model 
is found under extreme circumstances such as Nagaoka 
ferromagnetism\cite{Nagaoka_PhysRev.147.392} in the limit of infinite interaction strength in the presence of a single hole or ferromagnetism in  
special lattice geometries \cite{Lieb_PhysRevLett.62.1201}, or lattices with flat band 
dispersions\cite{Tasaki_PhysRevLett.69.1608,cmp/1104254245} etc. Alternatively, it is
also shown to occur for lattices with frustration where the density of states is asymmetric with the spectral weight shifted to the lower energy band 
edge\cite{Ulmke_EPJB1998,Wegner_PhysRevB.57.6211}. 
In multi-band systems with degenerate orbitals, the Hund's exchange coupling $J$ which 
favors intra-atomic ferromagnetic alignment comes into play and its role in 
stabilizing ferromagnetism has also been explored.
Since the interactions in the Hubbard model are purely local, the effects of
the kinetic energy of the electrons and hence the lattice structure also comes into 
question. Several studies have explored these issues in the past within the multi-band 
Hubbard model. However the multi-band problem is theoretically much harder 
and in most of the studies, the spin-flip and pair hopping processes in the intra-atomic exchange term are often dropped, and also the model is studied in the limit of 
infinite lattice dimensions.
Nevertheless, for the two-band Hubbard model with one electron per site, an effective Hamiltonian in the strong coupling was shown to describe a ferromagnetic (FM) order coexisting with a staggered orbital order the ground state\cite{KugelKhomskii,CyrotLyon}.
DMFT studies of the two-band Hubbard model in the 
limit of infinite lattice dimensions found metallic ferromagnetism
for large $J$ and at band fillings with more than one electron per site 
(i.e.~$N>1$), where it can be explained by the double
exchange mechanism\cite{HeldVollhardt_EPJB,Momoi_PhysRevB.58.R567,
Pruschke_PhysRevB.81.035112,Peters_PhysRevB.83.125110}. 
At quarter-filling ($N=1$), FM order was found to coexist with a staggered orbital order 
but in the insulating phase\cite{Momoi_PhysRevB.58.R567,Peters_PhysRevB.83.125110}.
SBMFT calculations found FM order in the two-band Hubbard
model for large $J$ where the FM phase preempts the Mott 
transition\cite{Fresard_PhysRevB.56.12909}. As opposed to an infinite lattice,
studies on square lattice using variational Monte Carlo based
on Jastrow type wave functions also support the above 
results\cite{Kubo_PhysRevB.79.020407,Becca_PhysRevB.98.075117,
Kubo_PhysRevB.103.085118} and in addition, it finds that when FM phase occurs at 
quarter-filling along with orbital order, it is always in the fully polarized 
state\cite{Becca_PhysRevB.98.075117}. 

The slave-rotor mean field (SRMF) theory was introduced to overcome some of the limitations of DMFT and yet as
a computationally inexpensive method, and was used to study Mott transition in the
multi-orbital Hubbard model in great details\cite{Florens_PhysRevB.70.035114}. 
However, by construction SRMF theory is unable to take
into account magnetic order as it introduces a single slave variable corresponding to
all the orbital degrees of freedom in a lattice site. In contrast  
the slave-spin mean-field (SSMF) theory\cite{Medici_PhysRevB.72.205124,
Qimiao_PhysRevB.86.085104,CaponeBookChapter} introduces an auxiliary slave variable  
for each of the spin-orbital indices in a site. Thus in principle, it is capable
of giving a symmetry broken solution where the mean field parameters depend upon
the orbital and spin indices. However, it turns out not true and as elaborated by
Georgescu {\it et al}~\cite{IsmailBeigiPhysRevB.96.165135}, the SSMF scheme does not break any symmetry and always gives a paramagnetic solution. To get around the problem, 
they introduced an approach based on total energy which is essentially a variational 
approach where one considers a symmetry broken variational wave function in the 
slave-particle representation and optimizes the energy with respect to the
parameters. The method which we term here as the variational slave-spin mean field (VSSMF) 
method is described in the following in detail.

In this work, we revisit the ground state phase diagram of the two-band Hubbard model 
as a function of Hubbard repulsion $U$ and Hund's exchange coupling $J$ using the VSSMF 
method. We show that at half-filling, the ground state evolves from a Slater
antiferromagnetic state with local charge fluctuations at smaller $U$ to an 
antiferromagnetic Mott insulating state with complete charge localization at large $U$ 
via a first order transition. At quarter-filling, the ground state remains paramagnetic
metallic at smaller $U$. At larger $U$ and in presence of Hund's coupling $J$, 
ferromagnetic order sets in with full spin polarization coexisting with an 
antiferro-orbital order. The system eventually becomes Mott insulating as $U$ is 
increased further. The rest of the paper is organized as follows. 
In Sec.~\ref{sec:model}, we describe the model and in Sec.~\ref{sec:method}, we 
describe the variational slave-spin mean-field method. The results are described
in Sec.~\ref{sec:results} and the final conclusion in Sec.~\ref{sec:conclusion}.   

\section{Model}
\label{sec:model}
We consider the following two-band Hubbard model on a square lattice,
\begin{align}
{\cal H} =& -t\sum_{\la i,j\ra m\sigma} \left(c^\dag_{im\sigma}c_{jm\sigma}+hc\right) 
+ U\sum_{im}n_{im\up}n_{im\dn} \nnum\\
&+ U'\sum_{im\neq m'} n_{im\up}n_{im'\dn}
+(U'-J)\sum_{im<m',\sigma}n_{im\sigma}n_{im'\sigma} \nnum\\
& -J\sum_{im\neq m'}c^\dag_{im\up}c_{im\dn}c^\dag_{im'\dn}c_{im'\up}
+J\sum_{im\neq m'}c^\dag_{im\up}c^\dag_{im\dn}c_{im'\dn}c_{im'\up}
\label{eq:hubbard3B}
\end{align}
where $c^\dag_{im\sigma}$ creates an electron at site $i$, orbital $m$ ($=1,2$) 
with spin $\sigma$, and $n_{im\sigma}=c^\dag_{im\sigma}c_{im\sigma}$.
The first term describes orbital diagonal hopping of electrons between nearest 
neighbor sites. The rest of the terms describe various interactions as follows.
The first three terms represent the Coulomb repulsion of electrons in the same 
orbital, in different orbitals with opposite spins, and in different orbitals with
the same spin. The last two terms are the spin-flip and pair hopping terms, respectively.
$U$ is intra-orbital interaction strength and $J$ is Hund's exchange coupling 
which favors atomic states with maximum total spin and orbital angular momenta. 
We take $U'=U-2J$ in which case the Hamiltonian becomes rotationally invariant with 
respect to both spin and orbital degrees of freedom. The above form of the 
Hamiltonian is relevant to $d$-orbital systems with cubic crystal field in which 
the five fold degenerate $d$ orbitals splits into a three fold degenerate $t_{2g}$ 
orbitals and two fold degenerate $e_{g}$ orbitals. Also in several transition
metal oxides has layered structure where interlayer couplings can
be neglected and it is sufficient to consider a two dimensional lattice. 

\section{Variational slave-spin mean field method}
\label{sec:method}
In the slave-spin formalism\cite{PhysRevB.72.205124,HassanPhysRevB.81.035106,
2016arXiv160708468D,PhysRevB.86.085104} the physical electron states are mapped 
to a fermionic quasiparticle states coupled to an auxiliary spin-$1/2$ degree of 
freedom. That is, for each site $i$ and spin-orbital index $\alpha\equiv m\sigma$, the electron states are written as
\begin{equation}
\begin{gathered}
|{n_{i\alpha}=0}\rangle \Rightarrow |{n^f_{i\alpha}=0, 
S^{z}_{i\alpha}=-1/2}\rangle \\
|{n_{i\alpha}=1}\rangle \Rightarrow |{n^f_{i\alpha}=1, 
S^{z}_{i\alpha}=+1/2}\rangle
\end{gathered}
\end{equation}
where $n_{i\alpha}=c^\dag_{i\alpha}c_{i\alpha}$ is the number operator for 
electrons and $n^f_{i\alpha}=f^\dag_{i\alpha}f_{i\alpha}$ 
is that for the
fermionic quasiparticles (spinons). $S^z_{i\alpha}$ is the auxiliary 
spin-$1/2$ degree of freedom coupled to the spin-orbital. 
The mapping enlarges the local Hilbert space introducing unphysical states
$|{n^f_{i\alpha}=0, S^{z}_{i\alpha}=+1/2}\rangle$ and 
$|{n^f_{i\alpha}=1, S^{z}_{i\alpha}=-1/2}\rangle$. These are in principle 
eliminated by enforcing the constraint, 
\begin{align}
n^f_{i\alpha} = S^z_{i\alpha} + \frac{1}{2}
\label{eq:exact_constraint}
\end{align}
The electron operators are decomposed accordingly as 
$c_{i\alpha}=f_{i\alpha}O_{i\alpha}$ and 
$c^\dag_{i\alpha}=f^\dag_{i\alpha}O^\dag_{i\alpha}$. Several representations
for $O_{i\alpha}$ are possible which act in the physical Hilbert space identically
but give different solutions when the constraint is enforced approximately.
In the $Z_2$ representation\cite{PhysRevB.72.205124,HassanPhysRevB.81.035106} 
which we use here, 
\begin{align}
O_{i\alpha} = S^{-}_{i\alpha}+g_{i\alpha}S^{+}_{i\alpha}
\label{eq:Z2_representation}
\end{align}
The gauge factors $g_{i\alpha}$-s are fixed by requiring that 
in the non-interacting limit, the solution matches with that of the corresponding original Hamiltonian. In this new representation, the kinetic energy term becomes,
\begin{align}
{\cal H}_t \equiv& -t\sum_{\la i,j\ra \alpha} 
\left(O^\dag_{i\alpha}O_{j\alpha}f^\dag_{i\alpha}f_{j\alpha}+hc\right) 
-\mu\sum_{i\alpha}n^f_{i\alpha} \nnum\\
& -\sum_{i\alpha}h_{i\alpha}\left[n^f_{i\alpha}-(S^z_{i\alpha}+\frac{1}{2})\right]
\end{align}
where we have introduced a chemical potential term to control the particle number.
The last term which is zero because of the constraint, is introduced 
to enforce the constraints at the mean field level as mentioned later, with
the factors $h_{i\alpha}$-s acting as Lagrange multipliers.
The density-density terms in the interaction part of the Hamiltonian can be
written entirely in terms of the spin variables. Since 
$n_{i\alpha}\equiv n^f_{i\alpha}=S^z_{i\alpha}-\frac{1}{2}$ by the constraint, 
we can write
\begin{align}
n_{im\sigma}n_{im'\sigma'} = S^z_{im\sigma}S^z_{im'\sigma'} +
\frac{1}{2}\bigl(S^z_{im\sigma} +S^z_{im'\sigma'}\bigr) + \frac{1}{4}
\end{align}
The spin-flip and the pair hopping terms can not be represented entirely in terms of the 
slave-spin operators. In most of the studies, these two terms are dropped as their 
contribution is small. However these can be represented approximately by replacing
$c^\dag_{i\alpha}$ with $S^{+}_{i\alpha}$ 
which reproduces the energy spectrum in the atomic limit\cite{2016arXiv160708468D}. 
Here we use this approximate mapping for these two terms. Dropping the constants, 
the interaction part of the Hamiltonian becomes,
\small{
\begin{align}
{\cal H}^S_{int} \equiv&\; U\sum_{im}\overline{S^z_{im\up}S^z_{im\dn}} 
+ U'\sum_{im\neq m'}\overline{S^z_{im\up}S^z_{im'\dn}}
+(U'-J)\sum_{im<m',\sigma}\overline{S^z_{im\sigma}S^z_{im'\sigma}} \nnum\\ 
&-J\sum_{im\neq m'}S^{+}_{im\up}S^{-}_{im\dn}S^{+}_{im'\dn}S^{-}_{im'\up}
+J\sum_{im\neq m'}S^{+}_{im\up}S^{+}_{im\dn}S^{-}_{im'\dn}S^{-}_{im'\up}
\end{align}}
where we have defined $\overline{S^z_{i\alpha}S^z_{i\alpha'}}=
S^z_{i\alpha}S^z_{i\alpha'} + \bigl(S^z_{i\alpha}+S^z_{i\alpha'}\bigr)/2$.
The full Hamiltonian in the slave-spin representation becomes,
\begin{align}
{\cal H} =& -t\sum_{\la i,j\ra \alpha} 
\left(O^\dag_{i\alpha}O_{j\alpha}f^\dag_{i\alpha}f_{j\alpha}+hc\right) 
-\mu\sum_{i\alpha}n^f_{i\alpha} \nnum\\
& -\sum_{i\alpha}h_{i\alpha}\left[n^f_{i\alpha}-(S^z_{i\alpha}+\frac{1}{2}) 
\right] + {\cal H}^S_{int}
\label{eq:ss_ham}
\end{align}
We take the ground state to be $\ket{\Psi} = \ket{\Psi_f}\ket{\Psi_S}$. 
The first component belongs to the spinon sector and is the ground state of the following spinon Hamiltonian,
\begin{align}
{\cal H}_f =& \bra{\Psi_S}{\cal H}\ket{\Psi_S} 
= -t\sum_{\la i,j\ra\alpha}\left(B_{i\alpha,j\alpha}f^\dag_{i\alpha}f_{j\alpha}
+hc\right)\nnum\\
& -\sum_{i\alpha}(\mu+h_{i\alpha})n^f_{i\alpha} 
\end{align}
Similary, the second component is the ground state of the slave-spin Hamiltonian,
\begin{align}
{\cal H}_S =& \bra{\Psi_f}{\cal H}\ket{\Psi_f} = 
-t\sum_{\la i,j\ra \alpha}\left(\chi_{i\alpha,j\alpha}O^\dag_{i\alpha}O_{j\alpha}
+hc\right)\nnum\\
& + \sum_{i\alpha}h_{i\alpha}\bigl(S^z_{i\alpha}+\frac{1}{2}\bigr) + {\cal H}^S_{int}
\end{align}
Thus we get two coupled Hamiltonians which need to be solved self-consistently. The
mean field parameters are given by,
\begin{align}
\chi_{i\alpha,j\beta} = \bra{\Psi_f} f^\dag_{i\alpha}f_{j\beta} \ket{\Psi_f},\quad
B_{i\alpha,j\beta} = \bra{\Psi_S} O^\dag_{i\alpha}O_{j\beta}\ket{\Psi_S}
\end{align}
The Lagrange multipliers $h_{i\alpha}$-s are adjusted so as to satisfy the constraint on the average,
\begin{align}
\bra{\Psi_f} n^f_{i\alpha} \ket{\Psi_f} = 
\bra{\Psi_S} S^z_{i\alpha} \ket{\Psi_S} + \frac{1}{2}
\end{align} 
The spinon Hamiltonian is a non-interacting one and can be solved readily.
The slave-spin Hamiltonian is fully interacting and we solve it using mean field.
In the single site approximation, we mean field decouple the terms connecting
the $i$-th site (cluster) to the other sites (bath) as,
\begin{align}
O^\dag_{i\alpha}O_{j\alpha} \approx O^\dag_{i\alpha}\Phi_\alpha,\quad 
j\in\text{bath} 
\end{align}
where $\Phi_{\alpha}=\la O_{j\alpha}\ra$ is the order parameter 
which is assumed independent of the site index due to translational symmetry.
The quasiparticle (QP) weight which plays a crucial role in the theory,
is defined as $Z_\alpha=|\Phi_\alpha|^2$.
It denotes the degree of charge fluctuation in the system and gives a 
measure of effective mass enhancement due to correlations, $m^*=m/Z_{\alpha}$. 
$Z_\alpha$ is unity in the non-interacting limit. As $U$ is increased, $Z_\alpha$ 
drops indicating a correlated metallic state. Mott transition is
indicated by the vanishing of $Z_{\alpha}$ at a critical interaction $U_{c}$.
In this single site approximation,
$B_{i\alpha,j\alpha}=Z_{\alpha}$. Therefore the spinon and slave-spin Hamiltonians
become,
\begin{align}
{\cal H}_f =& -t\sum_{\la i,j\ra\alpha}Z_{\alpha}\left(f^\dag_{i\alpha}f_{j\alpha}
+hc\right)-\sum_{i\alpha}(\mu+h_{i\alpha})n^f_{i\alpha}\label{eq:H_spinon} \\
{\cal H}^{site}_S &= \sum_{\alpha}\left(\eta_{\alpha}O^\dag_{i\alpha} + hc\right)
 + \sum_{\alpha}h_{i\alpha}\bigl(S^z_{i\alpha}+\frac{1}{2}\bigr) + {\cal H}^S_{int}
 \label{eq:H_slave}
\end{align}
where $\eta_{\alpha} = -t\Phi_\alpha\sum_{j}\chi_{i\alpha,j\alpha}$ with the sum
being over the neighboring sites of $i$. The gauge factors in 
Eq.~(\ref{eq:Z2_representation}) are obtained by requiring $Z_{\alpha}=1$ in the
limit $U=J=0$ which gives\cite{2016arXiv160708468D,HassanPhysRevB.81.035106}
\begin{align}
g_{i\alpha} = \frac{1}{\sqrt{n^0_\alpha(n^0_\alpha-1)}}-1,\quad
n^0_\alpha = \bra{\Psi^0_f} n^f_{i\alpha} \ket{\Psi^0_f}
\end{align}
where $\ket{\Psi^0_f}$ is the spinon ground state obtained by putting $U=J=0$ 
in the slave-spin Hamiltonian. The non-interacting solution yield the
Lagrange multipliers given by,
\begin{align}
h^0_{i\alpha} = -e^0_\alpha(1-n^0_\alpha)(1+g_{i\alpha})^2,\quad 
e^0_\alpha= \eta_\alpha/\Phi_\alpha
\end{align}
If non-zero, it introduces an extra chemical potential to the spinon 
Hamiltonian even in the non-interacting limit which is not desirable.
To nullify the effect, one needs to shift the chemical potential
$\mu$ to $\mu-h^0_{i\alpha}$ in Eq.~(\ref{eq:H_spinon}). With all the
ingredients in place, Eq.~(\ref{eq:H_spinon}) and (\ref{eq:H_slave}) are 
solved self-consistently to find all the mean field parameters and the 
quasi-particle weight $Z_{\alpha}$. Though in principle, these parameters 
depend upon the index $\alpha$ and can give rise to spin-resolved solutions, 
in practice the obtained solutions always turn out to be 
non-magnetic\cite{IsmailBeigiPhysRevB.96.165135}. This is at first 
disappointing given that the whole formalism is based on introducing 
an auxiliary variable for each spin-orbital. However, symmetry broken solutions
can be obtained by introducing an external field and minimizing the total energy
as described below. 

\subsection{Variational formalism}
In the variational formalism, we introduce symmetry breaking fields $b_{i\alpha}$
to the spinon Hamiltonian and write
\begin{align}
{\cal H}_f^{var} =& -t\sum_{\la i,j\ra\alpha}Z_{\alpha}\left(f^\dag_{i\alpha}f_{j\alpha}
+hc\right)-\sum_{i\alpha}(\mu+h_{i\alpha})n^f_{i\alpha}  \nnum\\
& + \sum_{i\alpha}b_{i\alpha}n^f_{i\alpha}
\label{eq:Hvar_spinon} 
\end{align}
The quantities $\{b_{i\alpha}\}$ constitute the variational parameters.
Let $|\Psi_f^{var}\rangle$ be the ground state of ${\cal H}_f^{var}$. With 
this, we take the following as the variational wave function of the 
slave-spinon Hamiltonian defined in Eq.~(\ref{eq:ss_ham}),
\begin{align}
\ket{\Psi_{var}(\{b_{i\alpha}\})} = \bigl|{\Psi_f^{var}(\{b_{i\alpha}\})}\bigr\ra
\bigl|{\Psi_S^{var}(\{b_{i\alpha}\})}\bigr\ra
\end{align}
Assuming that the wave function is normalized, the variational energy is 
obtained as
\begin{align}
E_{var}&(\{b_{i\alpha}\}) = \bra{\Psi_{var}}{\cal H}\ket{\Psi_{var}} \nnum\\
=& -t\sum_{\la i,j\ra \alpha}
\left\la\left(O^\dag_{i\alpha}O_{j\alpha}f^\dag_{i\alpha}f_{j\alpha}+hc\right)\right\ra 
-\sum_{i\alpha}(\mu+h_{\alpha})\left\la n^f_{i\alpha}\right\ra \nnum\\
& +\sum_{i\alpha}h_{i\alpha}\Bigl\la(S^z_{i\alpha}+\frac{1}{2})\Bigr\ra 
+ \left\la{\cal H}^S_{int}\right\ra
\end{align}
In the single site approximation, the above expression for the energy becomes,
\begin{align}
E_{var} =& \sum_{\alpha}e_{\alpha} -\sum_{i\alpha}(\mu+h_{\alpha})\la n^f_{i\alpha}\ra_f 
+\sum_{i\alpha}h_{i\alpha}\bigl\la(S^z_{i\alpha}+\frac{1}{2})\bigr\ra_S \nnum\\
& +\la{\cal H}^S_{int}\ra_S 
\end{align}
where $e_{\alpha}=-t\sum_j Z_{\alpha}\chi_{i\alpha,j\alpha}$ and the sum is over the 
the nearest neighbor sites of $i$. In the above the averages are calculated 
with respect to the variational states.
The variational wave function $|{\Psi_{var}(\{b_{i\alpha}\})}\rangle$ is determined by
setting the external fields as in Eq.~(\ref{eq:Hvar_spinon}) and
solving ${\cal H}_f^{var}$ and ${\cal H}_S$ self-consistently.
The variational energy is minimized with respect to the external fields and
the optimal solution determines the variational ground state.

\section{Results}
\label{sec:results}
We discuss the results of our calculations for the ground state of the two-band Hubbard model 
at two integer fillings, e.g.~at $N=2$ (half-filling) and $N=1$ (quarter-filling).  
In the variational calculations, at half-filling, we consider a single external
field corresponding to a two-sublattice antiferromagnetic order given by
$b_{im\sigma}=(-1)^i \eta(\sigma)b_{AF}$, where the $\eta(\sigma)$ is $+$ ($-$) for 
$\sigma = \up$ ($\dn$). 
At quarter-filling, we simultaneously apply two fields. 
One corresponds to ferromagnetic order, $b_{im\sigma}=-\eta(\sigma)b_{FM}$ and 
the other corresponds to an antiferromagnetic orbital order, 
$b_{im\sigma}=(-1)^{m+i} b_{l}$ where $m=1,2$ and $i$ is the site index. 
In the followings, the energy values mentioned are energy per site in units of $t$.

\subsection{Half-filling}
\begin{figure}[!htb]
 \centering
  \includegraphics[width=0.7\columnwidth]{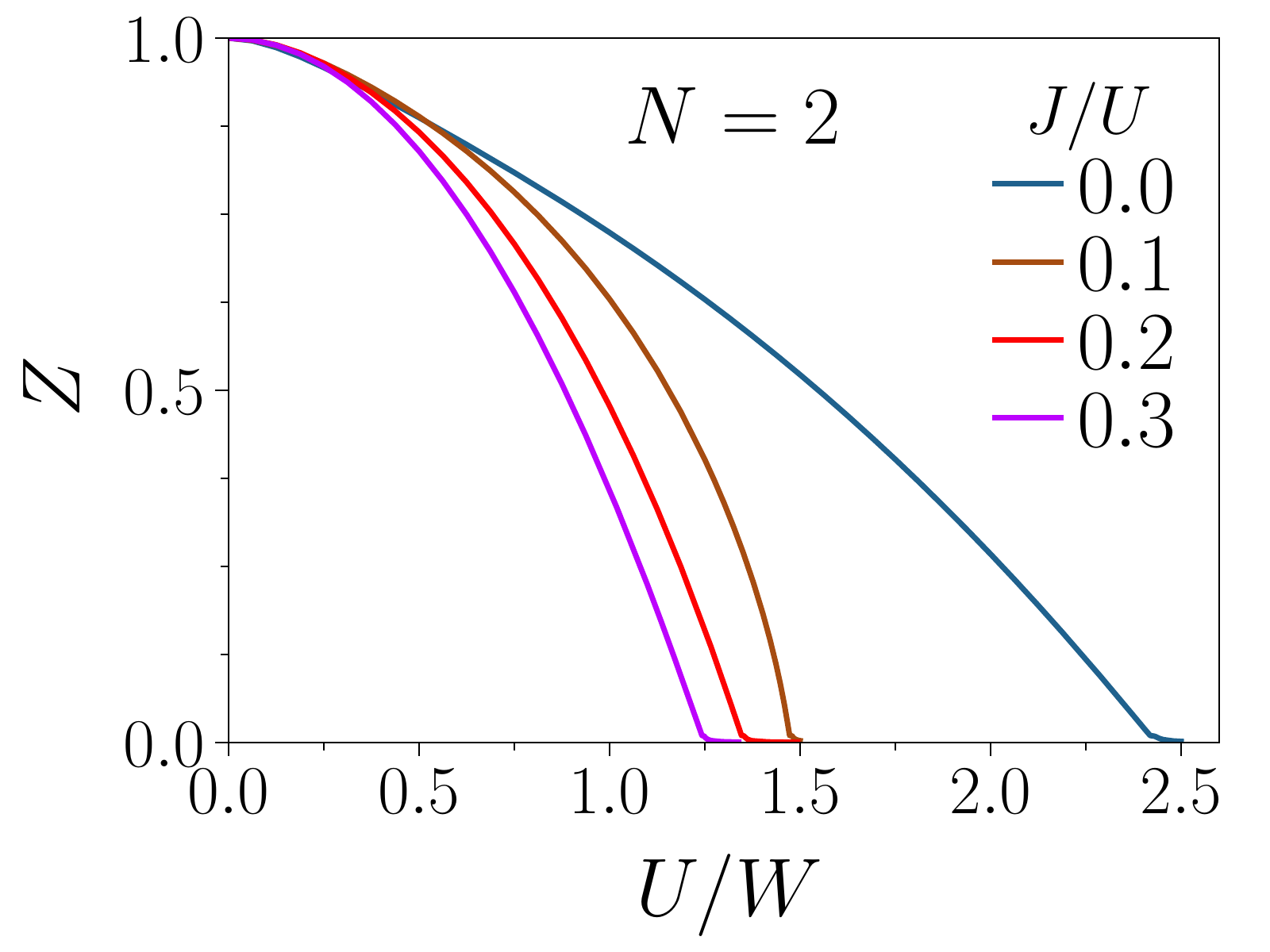}
  \caption{Mott transition in the two-band Hubbard model at half-filling in the
  paramagnetic sector and the effect of Hund's coupling $J$. 
  The figure shows the QP weight $Z$ as a function of $U/W$ at various Hund's 
  couplings $J/U$. At half-filling, $J$ reduces the critical interaction $U_c$ drastically.}
  \label{fig:01}
\end{figure}

Here we discuss the results for $N=2$ particles per site, that is the half-filled
band. First, let us consider the paramagnetic sector. The quasiparticle 
weight $Z$ in the PM sector are shown in Fig.~\ref{fig:01} 
as a function of $U/W$ where $W$ is the non-interacting bandwidth, at various Hund's 
couplings $J/U$. It agrees well with results from previous SSMF 
studies\cite{CaponeBookChapter}.
$Z$ is unity in the non-interacting case at $U=0$. It decreases with increasing $U$ 
signifying a correlated metallic state for small $U$. As $U$ is further
increased, $Z$ vanishes continuously at a critical interaction strength $U_c$ beyond 
which we get the Mott insulating state. 
The $U_c$ value decreases sharply with 
increasing $J$. This $J$ dependence of $U_c$ can be qualitatively understood as follows. 
The atomic Mott gap for a charge transfer between two isolated atoms with $N=2$ electrons
per atom is $\Delta_{at}=U+J$\cite{Georges_AnnRevCondMat}. 
A condition for Mott transition is obtained by
equating $\Delta_{at}$ to the effective bandwidth $\bar{W}$ in presence of $J$ which
gives $U_c\approx \bar{W}(J)-J$. Thus $U_c$ decreases with $J$ as it correlates
the system more. The decrease is not linear in $J$ as the effective bandwidth 
$\bar{W}$ is also reduced by $J$ because of its effect on the intra-atomic ferromagnetic alignment of the spins. 

\begin{figure}[!htb]
 \centering
  \includegraphics[width=0.9\columnwidth]{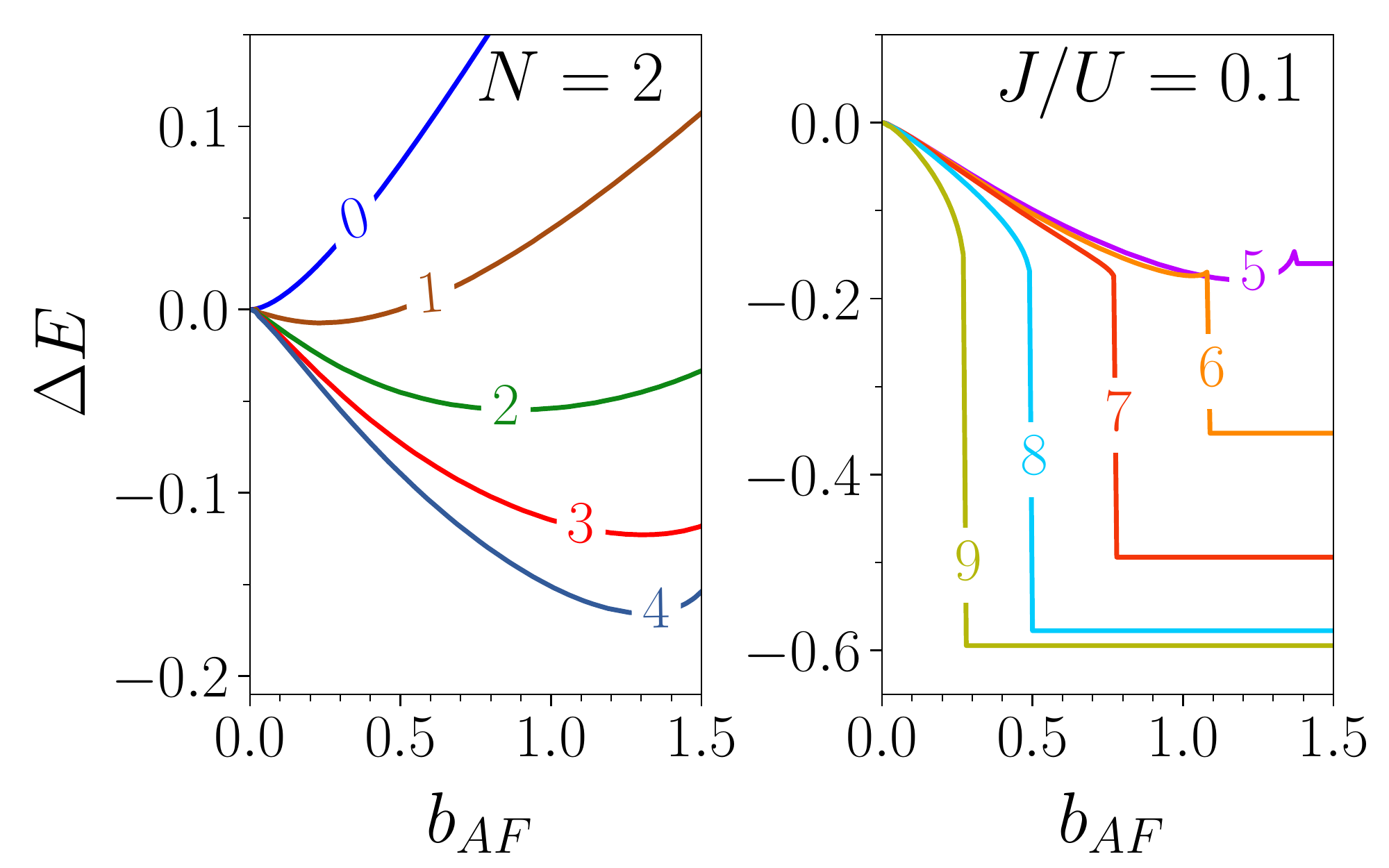}
  \caption{Energy gain $\Delta E=E_{var}(b_{AF})-E_0$ of the AF state as a function of 
$b_{AF}$ for different values of $U/t$ shown on the curves.
$J/U=0.1$, $N=2$.}
  \label{fig:02}
\end{figure}

\begin{figure}[!htb]
\centering
\includegraphics[width=0.7\columnwidth]{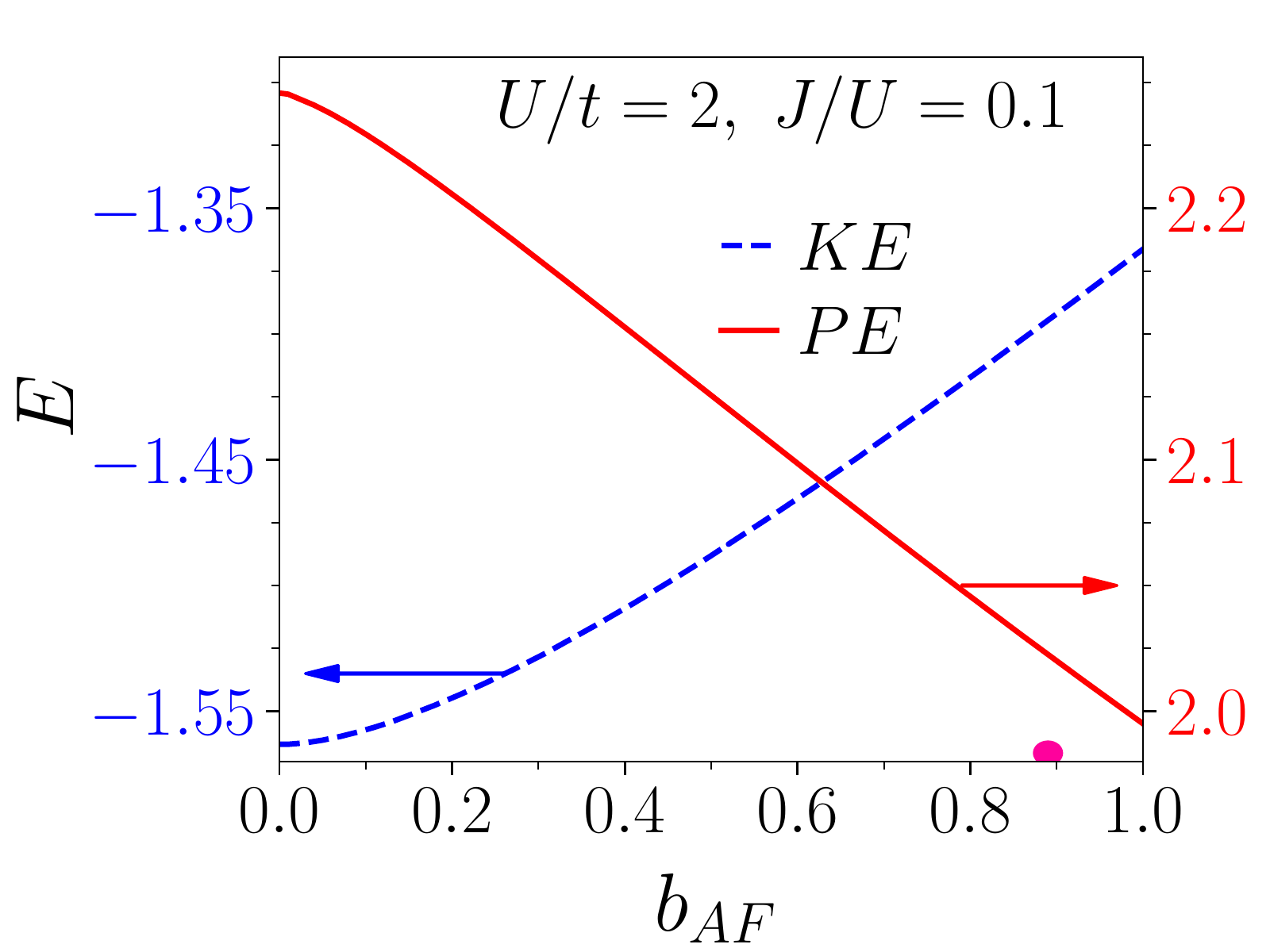}
\caption{Kinetic energy ($KE$) and potential energy ($PE$) as a function of
$b_{AF}$ at $U/t=2$, $J/U=0.1$. The pink dot near the $x$-axis indicates 
the point of minimum total energy. Particles per site, $N=2$.} 
\label{fig:03}
\end{figure}

\begin{figure}[!htb]
\centering
\subfigure[\label{fig:04a}]{
\includegraphics[width=0.7\columnwidth]{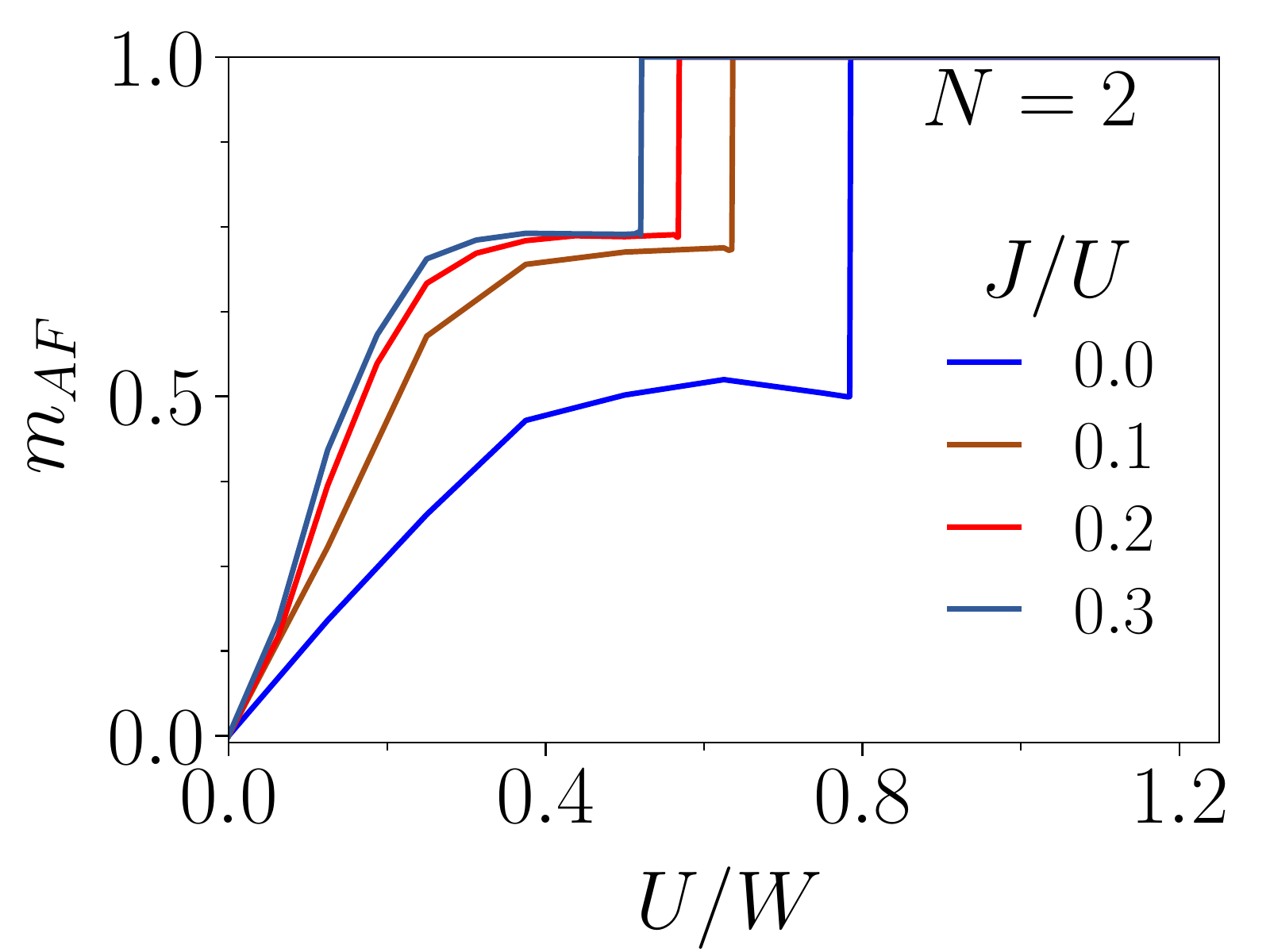}}
\subfigure[\label{fig:04b}]{
\includegraphics[width=0.7\columnwidth]{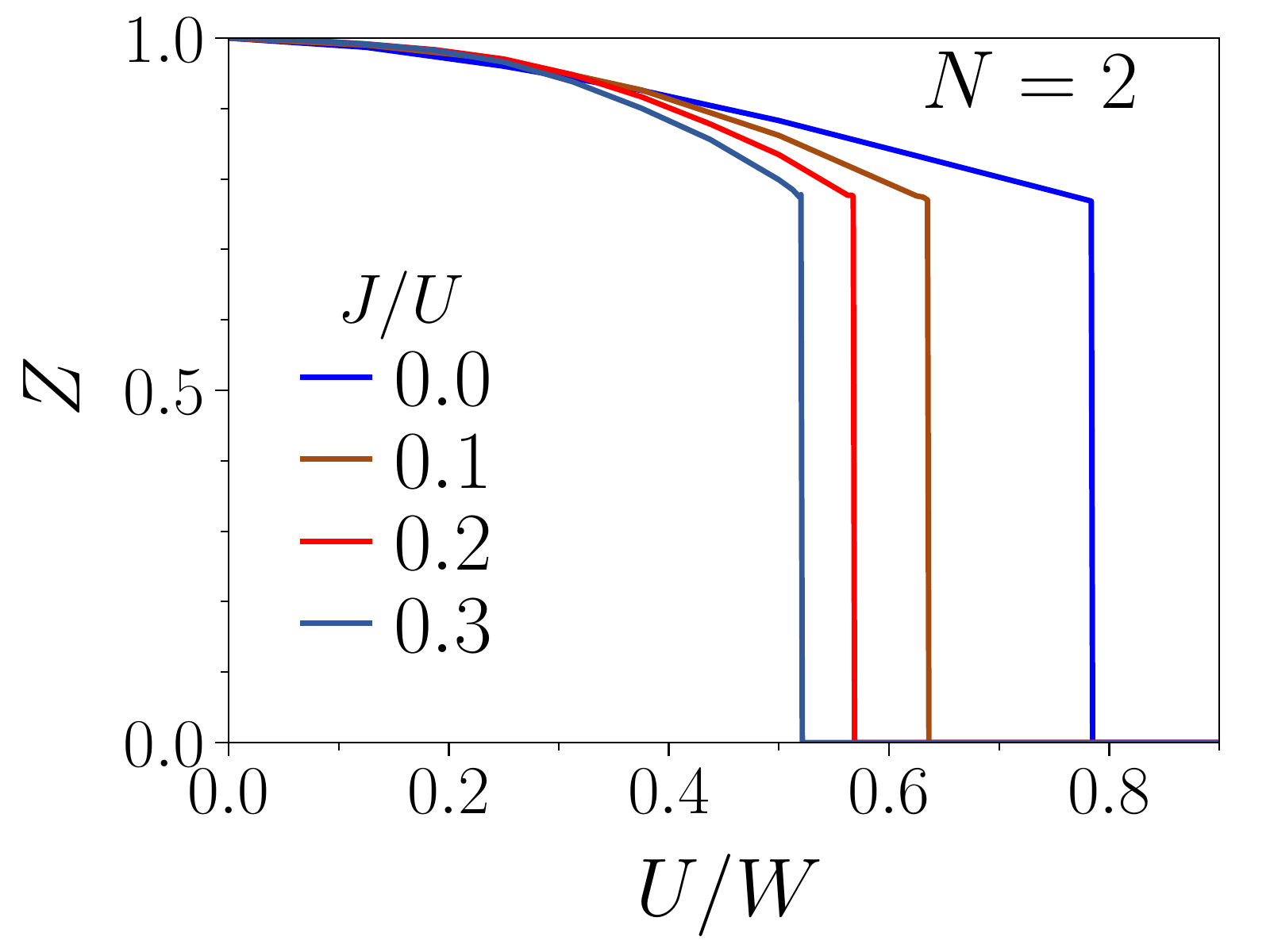}}
\caption{(a) AF order parameter $m_{AF}$ and (b) quasiparticle weight $Z$ as a function 
of $U/W$ at various $J/U$.}
\label{fig:04}
\end{figure}

Now we turn to the variational calculations. We apply an external symmetry breaking
field given by $b_{im\sigma}=(-1)^i\eta(\sigma)b_{AF}$ which favors a two 
sub-lattice AF order with the quantity $b_{AF}$ acting as the variational parameter.
We optimize the resulting variational wave function by minimizing the energy of the 
original Hamiltonian with respect to $b_{AF}$. The energy gain in the 
AF state is given by $\Delta E=E_{var}(b_{AF})-E_0$ where $E_0$ is the energy of the 
PM state. Fig.~\ref{fig:02} shows $\Delta E$ as a function of 
$b_{AF}$ for various $U$ at a fixed $J/U=0.1$.
As the figure shows, the minimum of the energy is obtained only at non-zero $b_{AF}$ for
all $U>0$. This means the ground state is antiferromagnetic at half-filling 
which is expected for the model with perfect Fermi surface nesting. 
For $U$ above a critical value, the energy drops suddenly to a low minimum and remains 
flat as function of $b_{AF}$. This happens due to the occurrence of Mott transition 
as discussed in the following. This lower energy in the AF state compared 
to the PM state is achieved via 
reduction in charge fluctuations which leads to less double occupancy and hence a 
gain in potential energy. This is shown in Fig.~\ref{fig:03} which plots the 
kinetic energy (KE) and potential energy (PE) components separately as a function 
of $b_{AF}$ for one particular $U$ and $J$. 
There is a loss in KE and gain in PE as $b_{AF}$ is increased and the total energy 
becomes minimum at a non-zero $b_{AF}$. We repeat the above optimization of the slave-spinon 
wave function at various points in the the parameter space and calculate  
the AF order parameter $m_{AF}=\frac{1}{L}\sum_i(-1)^i\la S^z_i\ra$ and the QP weight $Z$ at each of the optimized solutions. The results for $m_{AF}$ and $Z$ are shown in  
Fig.~\ref{fig:04}. 
As the figure shows, the scenario in this case is strikingly different to the 
results in the PM sector. The system develops AF ordering as soon as the interaction $U$ 
is turned on. The AF order reduces translational symmetry and doubles 
the unit cell which leads to the folding of the two degenerate bands. Therefore
the system becomes a Slater AF insulator. There are however fairly large local charge 
fluctuations as the value of $Z$ shows (Fig.~\ref{fig:04b}), though 
the state is globally incompressible. The AF order parameter $m_{AF}$ 
increases linearly with $U$ for small $U$. This is driven by the gain in potential energy as
double occupancies are reduced. However for larger $U$, it reaches a peak and starts to 
drop. This happens due to the system approaching the Heisenberg limit where the
AF gap is dominated by the superexchange process which goes as $\sim -4t^2/U$. 
In fact, it was shown that the Slater and the
Heisenberg limits are separated by the Mott transition where the gains in KE and PE
components switch sign\cite{Tremblay_PhysRevB.95.235109}. 
We do not see this in our results, likely because 
the single site approximation used here to solve the interacting slave-spin problem 
does not take into account spatial correlations properly.
Increasing Hund's coupling $J$ favors further localization and gives higher $m_{AF}$  
driven by lower onsite energy when both the spins in the two orbitals are parallel.
The QP weight $Z$ decreases with increasing $U$ and vanishes abruptly at a critical interaction
strength $U_c$ where $m_{AF}$ jumps to saturation giving rise to the N\'{e}el 
order. Thus we get a first order
transition from a Slater AF state with local charge fluctuations to an AF Mott 
insulating state with localized charge degree of freedom. Indeed within the single site 
formalism as mentioned above, the system goes to the atomic limit where the kinetic 
energy vanishes completely. One notable feature of the results here is that
the paramagnetic sector underestimates the effect of correlation and the Mott transition
occurs at much higher interaction. In the AF state, the $U_c$ values are much lower. 
For instance, at $J=0$, $U_c\sim W$ in AF state whereas it is $\sim 2.5 W$ in the 
PM state. Introducing $J$ further reduces $U_c$ by correlating the system 
more as mentioned before.

\subsection{Quarter-filling}
The case of quarter-filling with $N=1$ particle per site is more interesting.
Previous studies of the two-band Hubbard model found itinerant
ferromagnetism at band fillings greater than one and for large 
$J$\cite{HeldVollhardt_EPJB, Momoi_PhysRevB.58.R567, Pruschke_PhysRevB.81.035112,
Peters_PhysRevB.83.125110}. At $N=1$, the FM order was found to coexist with staggered 
orbital order but in an insulating 
phase\cite{Momoi_PhysRevB.58.R567,Peters_PhysRevB.83.125110}. 
Variational Monte Carlo studies of the model on 2D lattice
also find coexisting staggered orbital order with fully saturated 
ferromagnetism\cite{Kubo_PhysRevB.79.020407,Becca_PhysRevB.98.075117,Kubo_PhysRevB.103.085118}. 

\begin{figure}[!htb]
\includegraphics[width=0.7\columnwidth]{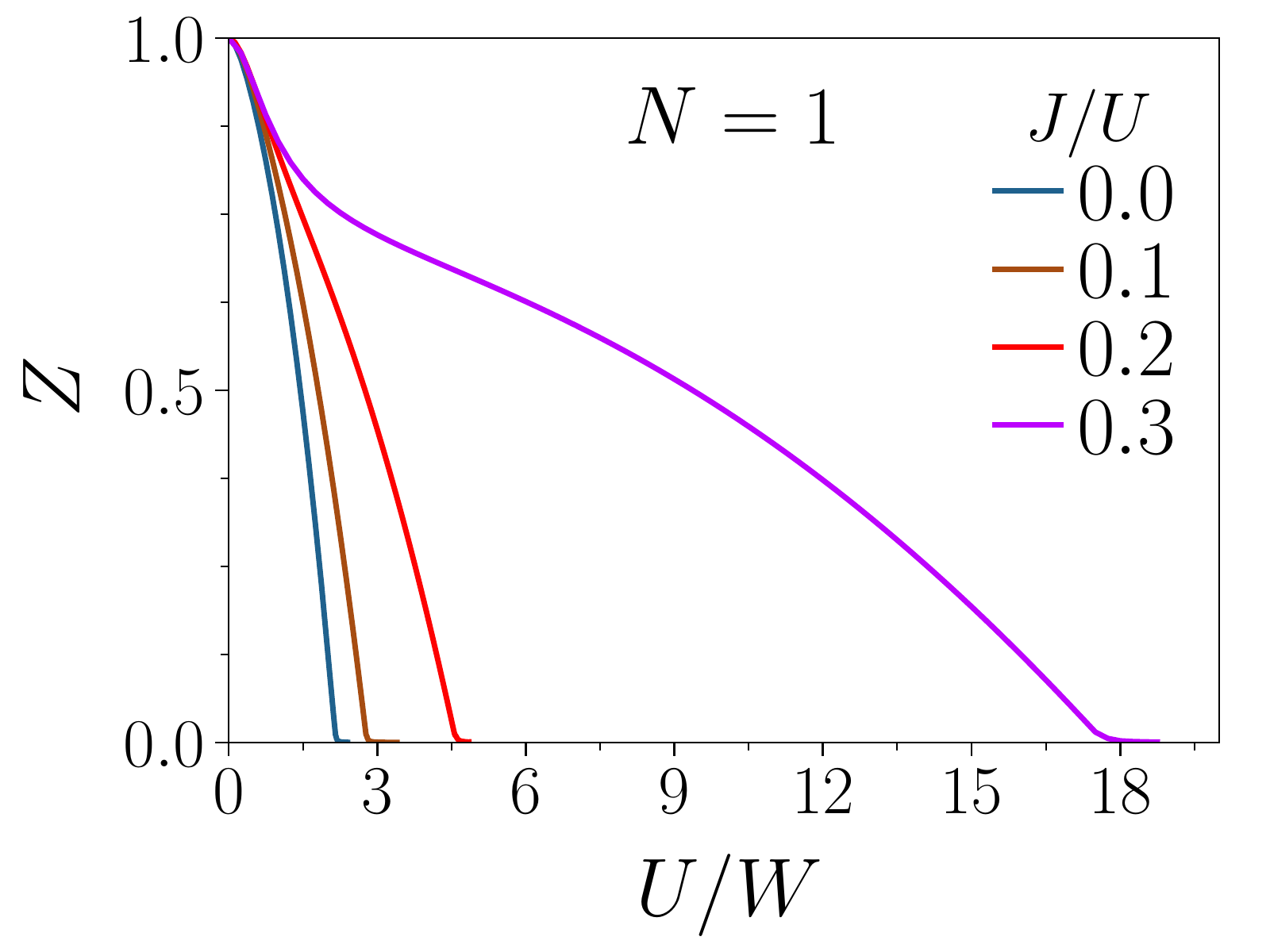}
\caption{Quasiparticle weight $Z$ as a function of $U/W$ at quarter-filling for 
various $J/U$.}
\label{fig:05} 
\end{figure}

\begin{figure}[!htb]
\includegraphics[width=0.8\columnwidth]{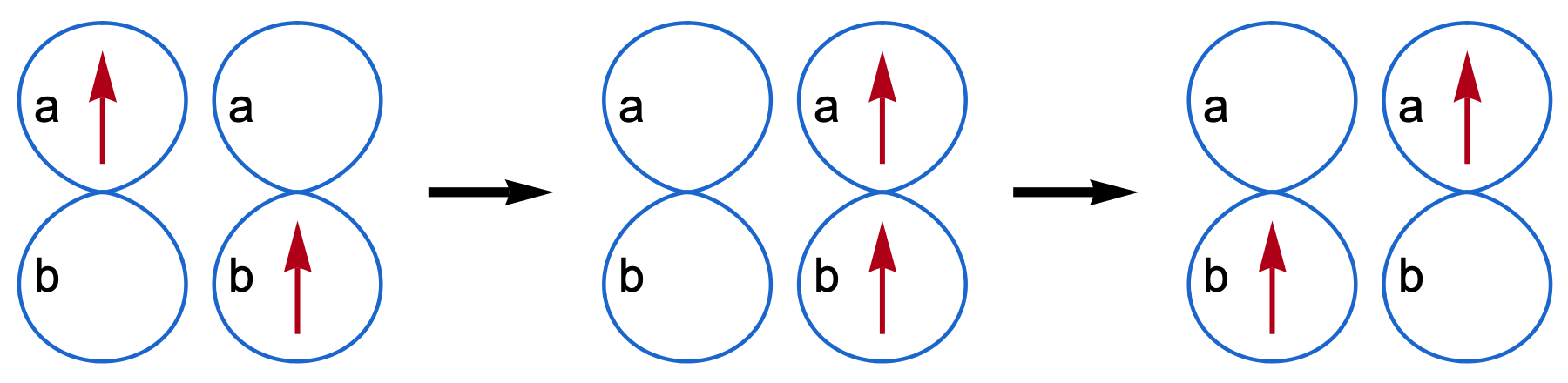}
\caption{The spin triplet and orbital singlet state of a two-site lattice with one electron per site. The kinetic exchange process shown schematically gives energy 
$\sim -4t^2/(U-3J)$ which is the lowest.}
\label{fig:06} 
\end{figure}

\begin{figure}[!htb]
\includegraphics[width=0.7\columnwidth]{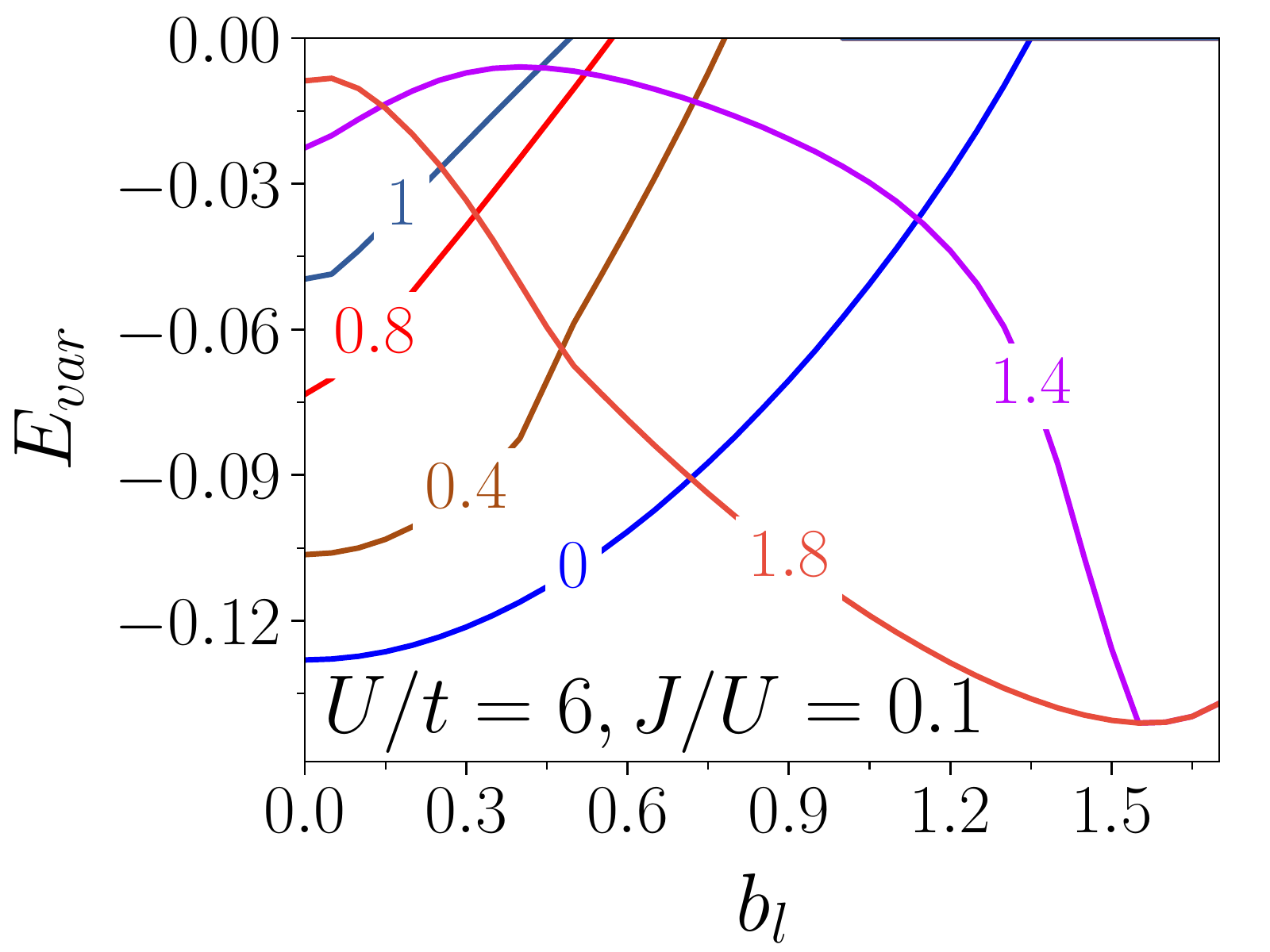}%
\caption{Energy $E_{var}$ versus $b_l$ for different values of $b_{FM}$ shown 
on the curves. $U/t=6$ and $J/U=0.1$. Particles per site, $N=1$.}
\label{fig:07}
\end{figure}

\begin{figure}[!htb]
\includegraphics[width=0.7\columnwidth]{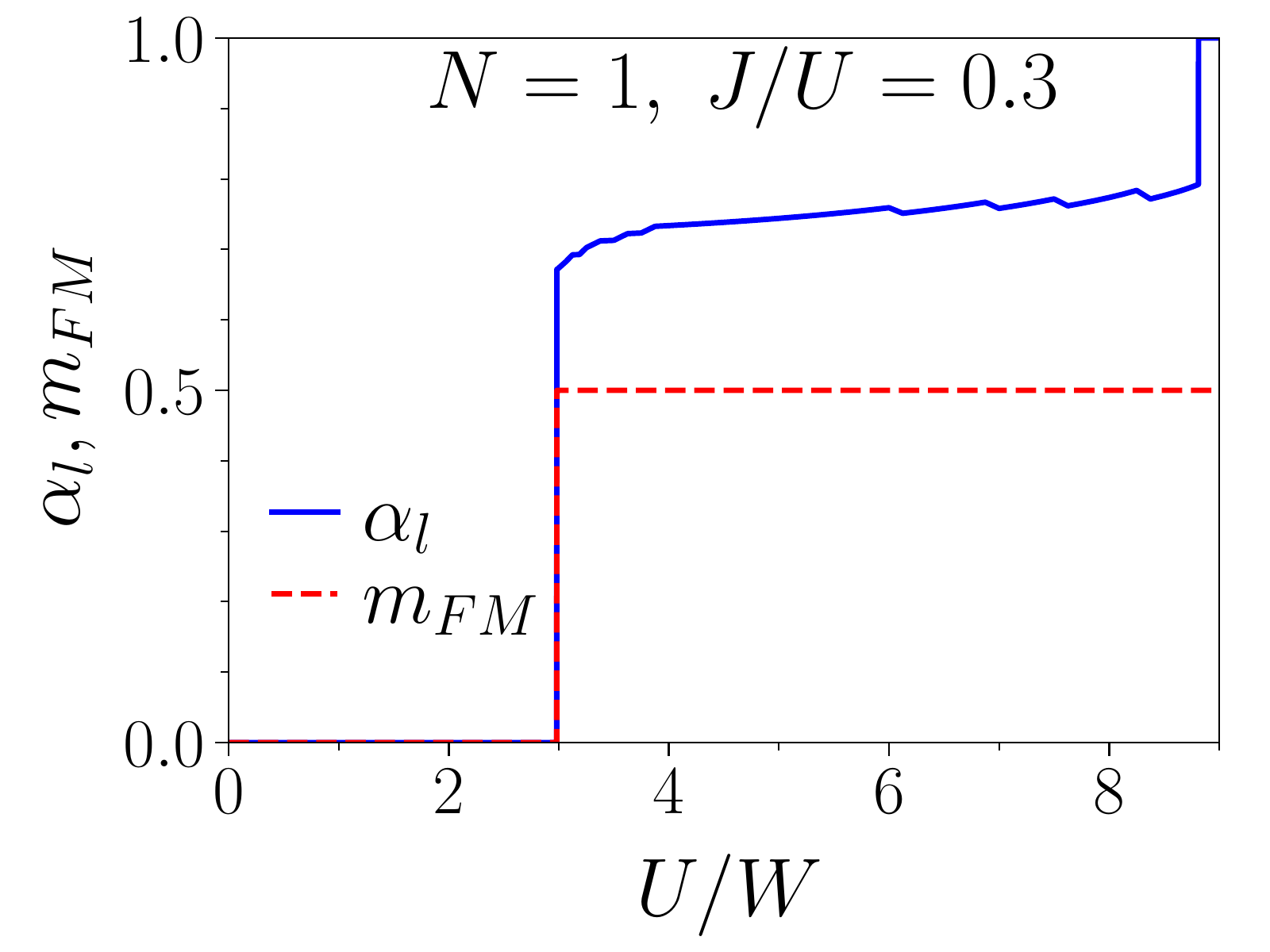}
\caption{Magnetic and orbital order parameters $m_{FM}$ and $\alpha_l$, respectively
as a function of $U/W$ at $J/U=0.3$.} 
\label{fig:08}
\end{figure}

\begin{figure}[!htb]
\includegraphics[width=0.95\columnwidth]{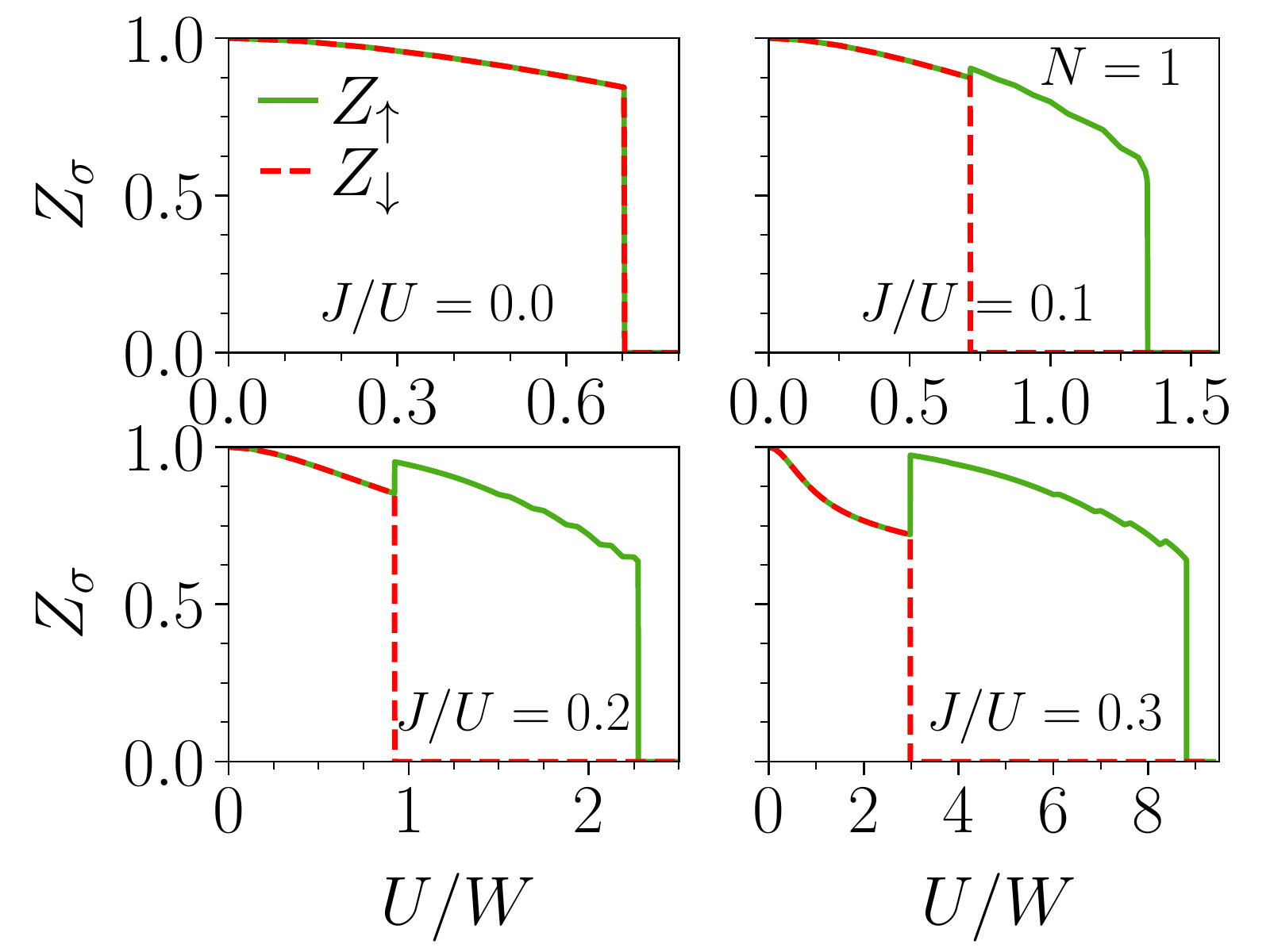}%
\caption{Quasiparticle weight $Z_{\sigma}$ versus $U/W$ for different $J/U$.}
\label{fig:09}
\end{figure}

Before we discuss the results from VSSMF calculations, it is again worthwhile to 
examine the paramagnetic sector results obtained using the simple SSMF method.
Fig.~\ref{fig:05} shows the QP weight in the paramagnetic sector 
as a function of interaction $U$. 
The system undergoes a continuous transition from paramagnetic metallic to Mott 
insulating state at a critical interaction $U_c$ which depends upon $J$. 
In contrast to the half-filled case, here the effect of $J$ on $U_c$ is the opposite. 
This can again be understood by looking at the atomic gap which at quarter-filling 
is given by $\Delta_{at} = U-3J$. Hence $U_c$ is estimated to vary as
$U_c\sim \bar{W}(J)+3J$. Thus increasing $J$ at this filling decorrelates the system more 
and increases $U_c$ substantially. Next, we consider the possibility of symmetry broken phase
using the variational approach. In the strong coupling limit and with $J>0$, one can easily see
that the ground state of a two-site lattice with one electron per site 
is a spin triplet and orbital singlet. This is shown schematically in 
Fig.~\ref{fig:06}. In this state, the sequence of hopping generated by the 
action of the low energy effective Hamiltonian creates an intermediate state of 
lowest energy thus making it the ground state of the system.
In order to examine the situation for the extended lattice, we apply two external 
fields, one given by $b_{im\sigma}=-\eta(\sigma) b_{FM}$ corresponding to the spin FM
order and the other by $b_{im\sigma}=(-1)^{m+i} b_{l}$ with $m=1,2$ being the orbital 
index, corresponding to the antiferro-orbital order. Thus here we have two variational
parameters, $b_{FM}$ and $b_l$. The calculations in this case become numerically more
extensive as the energy needs to be optimized with respect to two variational 
parameters at each point in the 
parameter space and each self-consistent calculation takes longer time to converge. 
We carried out the calculations and find that at smaller $U$, the energy minimum is
obtained only at $b_{FM}=b_l=0$ implying paramagnetic ground state at these values
of $U$. At $J=0$, the 
system continues to be paramagnetic metal as $U$ increased until MIT occurs at a higher $U$. 
However for non-zero $J$, as $U$ is increased above a lower critical value $U_{c1}$,
the energy gets substantially lowered by simultaneous application of 
fields $b_{FM}$ and $b_{l}$. This is shown in Fig~\ref{fig:07} where we plot 
the variational energy as a function of these two parameters for particular value of $U$ 
and $J$.
Thus for $U>U_{c1}$ we get a ferromagnetic state coexisting with antiferro-orbital 
order. If we look at the magnetization per site, we find it to be fully saturated with 
all the spins flipped parallel. This fully spin polarized state with orbital ordering
was also obtained in variational Monte Carlo studies\cite{Kubo_PhysRevB.79.020407,
Becca_PhysRevB.98.075117, Kubo_PhysRevB.103.085118}. The orbital ordering is however 
not saturated as it is disrupted by charge fluctuations. We calculate the 
staggered orbital order parameter as, 
$\alpha_l=\frac{1}{L}\sum_{i} (-1)^i\la n_{ia}-n_{ib}\ra$ where $n_{ia}$ ($n_{ib}$) is
the electron number in orbital $m=1$ ($m=2$) of a site `$i$'. 
The magnetization per site, $m_{FM}$ and $\alpha_l$ are shown in Fig.~\ref{fig:08}.
As $U$ is increased further, the system becomes Mott insulating via a first order 
transition at a higher critical interaction $U_{c2}$. In the Mott state, 
the charge degree of freedom gets frozen and the antiferro-orbital order 
becomes complete. The values of $Z_{\sigma}$ are shown in Fig.~\ref{fig:09} 
as a function of $U$ at various $J/U$.
In the paramagnetic state $Z_{\up}=Z_{\dn}$ and it decreases with increasing $U$.
In the case of finite Hund's coupling $J$, $Z_{\dn}$ drops to zero at $U=U_{c1}$ 
as the down spin carrier density vanishes at the transition. The QP weight for up-spin,
$Z_{\up}$ shows a jump at the transition driven by a gain in the kinetic energy
by the superexchange process as mentioned before. $Z_{\up}$ eventually drops to
zero at the higher critical interaction $U_{c2}$. Thus in this orbitally degenerate
system, the Hund's exchange coupling which favors intra-atomic spin alignment
in isolated atoms also brings about ferromagnetic order in the band limit.


\section{Conclusion}
\label{sec:conclusion}
We have revisited the ground state of the orbitally degenerate two-band Hubbard model on
a square lattice using variational slave-spin mean field theory. The ordinary 
slave-spin mean field theory which is a numerically less intensive technique to study
correlated electron systems, gives solutions where no symmetry is broken. This work extends 
the method to study symmetry broken phases thus making it more powerful. Using the method,
we find that at half-filling, the ground state of the model at smaller $U$ is a Slater antiferromagnetic insulator with sizeable local charge fluctuations. As $U$ is increased, 
the system goes to a Mott insulating state with N\'{e}el antiferromagnetic order via a first
order transition. The critical interaction $U_c$ for the Mott transition is much smaller 
compared to the corresponding value in the paramagnetic sector. Increasing Hund's coupling 
$J$ correlates the system more and reduces $U_c$. At quarter-filling, the ground state is paramagnetic metallic at smaller $U$. When Hund's coupling is present, the system goes to a 
fully spin-polarized ferromagnetic state with a coexisting antiferro-orbital order as
$U$ is increased above a critical value $U_{c1}$. 
The spinon band structure becomes gapped due to the staggered 
orbital ordering, but local charge fluctuation persists. As $U$ is increased further, 
a first order transition takes place at a critical interaction $U_{c2}$ where charge
fluctuation vanishes completely giving rise to a Mott insulating state.

\bibliography{references}
\end{document}